\documentclass[nofootinbib,aps,pra,twocolumn,superscriptaddress,floatfix]{revtex4-2}
\usepackage[english]{babel}
\usepackage[utf8]{inputenc}
\usepackage[T2A]{fontenc}
\usepackage{hyperref, xcolor, graphicx}
\usepackage{amsfonts, amssymb, amsmath, mathtools}
\usepackage{subcaption}
\usepackage{siunitx}
\usepackage{tikz}
\usepackage{calc}
\usepackage{float}
\usepackage{ragged2e}
\usepackage{soul}
\usepackage{dcolumn}
\usepackage[normalem]{ulem}
\usepackage{xcolor,colortbl}
\usepackage{array, makecell} 

\definecolor{Gray}{gray}{0.85}

\usepackage{bibunits}

\defaultbibliographystyle{unsrt}

\hypersetup{
    colorlinks=true,
    linkcolor=black,
    citecolor=blue,   
    urlcolor=blue,
}

\makeatletter
\def\maketitle{
\@author@finish
\title@column\titleblock@produce
\suppressfloats[t]}
\makeatother

\begin{document}

\title{
The prospects of nonthermal magnetization switching in near-compensated rare earth iron garnets 
}

\author{N.I.\ Gribova}
    \email{gribova.ni@phystech.edu}
    \affiliation{Moscow Institute of Physics and Technology, Dolgoprudny 141701, Russia}
\author{D.O.\ Ignatyeva}
    \affiliation{Russian Quantum Center, Moscow 121205, Russia}
    \affiliation{Lomonosov Moscow State University, Moscow 119991, Russia}
\author{N.A.\ Gusev}
    \affiliation{Russian Quantum Center, Moscow 121205, Russia}
\author{A.K.\ Zvezdin}
    \affiliation{Prokhorov General Physics Institute of the Russian Academy of Sciences, Moscow 119991, Russia}
    \affiliation{Russian Quantum Center, Moscow 121205, Russia}
\author{V.I.\ Belotelov}
    \email{belotelov@physics.msu.ru}
    \affiliation{Russian Quantum Center, Moscow 121205, Russia}
    \affiliation{Lomonosov Moscow State University, Moscow 119991, Russia}

\date{\today}

\begin{abstract}
Ultrafast spin dynamics in a magnetically compensated rare earth iron garnet film driven by femtosecond optical pulses through the inverse Faraday effect is theoretically investigated. Numerical simulations based on the equations of motion for the N\'eel vector reveal the temporal evolution of the system and its trajectories in the effective potential landscape tuned by external field and temperature. The results demonstrate a clear threshold behavior: weak pulses induce only oscillations around the initial equilibrium state, while a stronger excitation results in a deterministic magnetization switching. The switching threshold is determined by the magnetic state of the sample on its phase diagramme as well as on the laser pulse helicity. This mechanism demonstrates a non-thermal and even non-absorptive pathway towards optomagnonic logic and memory devices.
\end{abstract}

\maketitle

\begin{bibunit}

Optical control of magnetization is a central challenge in spintronics and ultrafast magnetism \cite{wang2021dual, kirilyuk2010ultrafast}. Femtosecond pulses enable spin reversal at picosecond timescales \cite{hennecke2025transient, stupakiewicz2017ultrafast, stupakiewicz2019selection, li2025picosecond, li2024field, hintermayr2024explaining, li2023timescales, yamada2025magnetization}, with significant perspective implications for high-speed data processing and quantum functionalities \cite{pal2024using, pirro2021advances, wang2022recent}. Research spans materials from metallic alloys to insulating garnets and synthetic multilayers \cite{mangin2014engineered, banerjee2020single, choi2025magnetization}.

Current optical switching (AOS) typically relies on light absorption, with the switching mechanism being either thermal or non-thermal --- {the} photoinduced magnetic anisotropy. In metallic ferrimagnets like GdFeCo \cite{ostler2012ultrafast, stanciu2007all}, laser-induced electron heating and circular dichroism drive helicity-dependent switching \cite{stanciu2007all, ostler2012ultrafast, vahaplar2009ultrafast} at fluences around 5 mJ/cm$^{-2}$ \cite{stanciu2007all, hennecke2025transient, ostler2012ultrafast, vahaplar2009ultrafast}. Alternatively, in dielectrics the coupling of light to spins does not {rely primarily on the} resonant heating of ions; in Co-doped iron garnets (YIG:Co) photo-excitation of localized Co d–d electronic transitions modifies the magnetic anisotropy and enables ultrafast magnetic switching \cite{stupakiewicz2019selection, zalewski2024ultrafast}.

In this work, we theoretically predict non-thermal switching in transparent ferrimagnets purely via the inverse Faraday effect (IFE) \cite{pitaevskii1961electric, van1965optically}. As a Raman-type nonlinear process, the IFE induces an effective magnetic field without absorption, significantly improving energy efficiency and operating speeds.

We propose utilizing transparent ferrimagnets near the magnetization compensation point $T_m$. The phase diagram includes collinear and non-collinear phases \cite{wu2024magnon, davydova2019h, krichevsky2023unconventional, ignatyeva2025high}. An important feature {of the non-collinear phase} is an existence of two degenerate equilibrium magnetization states~\cite{ignatyeva2025spin}. {Spin dynamics of such ferrimagnets is much richer compared to ferromagnetic materials, and includes excitation of the two modes with close frequencies}~\cite{krichevsky2023unconventional}, ferromagnetic mode softening and amplitude growth~\cite{ignatyeva2025high} and other peculiar effects \cite{blank2021thz,deb2018controlling, stupakiewicz2021ultrafast, ivanov2019ultrafast}. We stressed here on the necessity of identifying a proper magnetic phase state of the sample for the optical switching. The IFE-based mechanism might enable a deterministic magnetic bit writing and the development of optomagnonic logic elements for next-generation quantum and spintronic devices.

Let's consider a ferrimagnet with uniaxial anisotropy near the magnetization compensation point $T_m$ in the non-collinear phase with two degenerate magnetization states. A prototypical material is a rare-earth iron garnet $\mathrm{R}_3(\mathrm{Fe}\mathrm{Me})_5\mathrm{O}_{12}$ (RIG) of cubic symmetry. While ions of $\mathrm{R}^{3+}$ appear on dodecahedral sites, ions of $\mathrm{Fe}^{3+}$ and $\mathrm{Me}^{3+}$ are located on both octahedral and tetrahedral sites \cite{kaczmarek2024atomic}. In RIGs, magnetizations of $\mathrm{Fe}^{3+}$ ions on octahedral ($\mathbf{M}_\mathrm{a}$) and tetrahedral ($\mathbf{M}_\mathrm{d}$) sites couple nearly antiparallel, yielding a net iron magnetization $\mathbf{M}_{\mathrm{Fe}}=\mathbf{M}_\mathrm{d}+\mathbf{M}_\mathrm{a}$. The rare-earth ion $\mathrm{R}^{3+}$ can be either magnetic on nonmagnetic. For definiteness we consider here the latter case. If metal ions $\mathrm{Me}^{3+}$ primarily occupy the tetrahedral sublattice (e.g., $\mathrm{Me}=\mathrm{Ga}^{3+}, \mathrm{Al}^{3+}$) then at some temperature $T_m$, these contributions balance ($\mathbf{M}_\mathrm{d} = -\mathbf{M}_\mathrm{a}$), causing the net magnetization to vanish. The compensation temperature is tunable via concentration of the substituting metal \cite{dubs2025magnetically}.

The magnetic state is described by the Néel vector $\mathbf{L} = \mathbf{M}_{\mathrm{d}} - \mathbf{M}_{\mathrm{a}}$ and the total magnetization $\mathbf{M}_{\mathrm{Fe}}$ \cite{davydova2019h, schlauderer2019temporal, baierl2016nonlinear}. Statically, 
if the external magnetic field is applied in-plane then at the temperatures sufficiently deviating from the compensation point $T_m$ the sample rests in the collinear magnetic phase. At this, all spins are directed along the external magnetic field.
However, in the vicinity of $T_m$, a smallcanting of $\mathbf{M}_d$ and $\mathbf{M}_a$ appears and L deviates from the external magneitc field while keeping in the plane defined by the external magnetic field and the magnetc anisotropy axis.
Since in the 
non-collinear phase canting is small, less than $0.1$~deg \cite{krichevsky2023unconventional}, an approximation $\mathbf{L} \parallel \mathbf{M}_{\mathrm{a}} \parallel \mathbf{M}_{\mathrm{d}}$ holds (Fig.~\ref{fig:scheme}). Defining $\mathbf{L}$ in spherical coordinates as $\mathbf{L} = L(\cos\theta\cos\phi, \cos\theta\sin\phi, \sin\theta)$ and $\mathbf{M}_{i}=M_{{i}}(\cos\theta_{{i}}\cos\phi_{{i}}, \cos\theta_{{i}}\sin\phi_{{i}}, \sin\theta_{{i}})$ with $i = \mathrm{d, a}$ (Supplementary), we consider equilibrium positions where $\mathbf{L}$ aligns near the $x$-axis (the external field direction).

\begin{figure}[h]       
\includegraphics[width=0.7\linewidth]{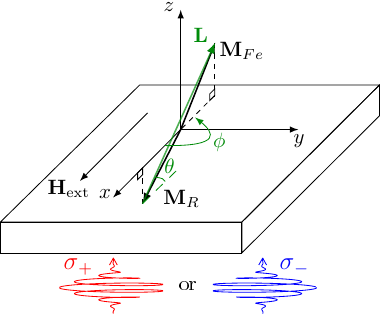}
\caption{\justifying{Ferrimagnetic film with uniaxial magnetic anisotropy in the external magnetic field $H_{\mathrm{ext}}$ along $x$ axis. The easy axis of the film is along $z$ axis. The $\theta$ angle is between $\mathbf{L}$ and XOY plane, $\phi$ is the angle between the projection $\mathbf{L}$ on the XOY plane and $x$ axis. Thin film is illuminated by circular polarized pulse $\sigma_+$ or $\sigma_-$.}}
	\label{fig:scheme}
\end{figure}

The equilibrium states of the two-sublattice system correspond to the minima of the effective potential energy of the system (Fig.~\ref{figure1}), which is obtained using the known relationship between the Lagrangian and the Hamiltonian ~\cite{davydova2019ultrafast, zvezdin1979dynamics, blank2021thz,krichevsky2023unconventional} (Supplementary) for the stationary case $\dot{\phi}=0$ and $\dot{\theta}=0$ \cite{krichevsky2023unconventional}:
\begin{align}
U_{\mathrm{eff}} = -\frac{\chi}{2} &H_{\mathrm{ext}}^2 (\sin^2\theta \cos^2\phi+\sin^2\phi) \\
&- m H_{\mathrm{ext}}\cos\theta\cos\phi - K\sin^2\theta,\nonumber
\end{align}
where $\chi$ is the transverse magnetic susceptibility {defined by the exchange energy}, $H_{\mathrm{ext}}$ is the external magnetic field, $m=M_\mathrm{d}-M_\mathrm{a}$ is the difference between the magnetization moduli of the sublattices, $K$ is the effective magnetic {uniaxial} anisotropy constant. The direction of $\mathbf{L}$ in equilibrium is determined by
\begin{equation}
\cos \theta_0 = \frac{m H_\mathrm{ext}}{ 2K+\chi H_\mathrm{ext}^2 },
\label{eq:theta0}
\end{equation}
{and $\phi_0=0$.}
Eq.~\eqref{eq:theta0} has two solutions, corresponding to two equilibrium states with $+\theta_0$ and $-\theta_0$ (Fig.~\ref{figure1}). These states are degenerate, as they have the same potential energy $U_{\mathrm{eff}}$. The height of the barrier between the equilibrium positions is 
\begin{equation}
    \Delta U_{\mathrm{eff}}=-mH_{\mathrm{ext}}+\frac{1}{2}\bigl[\chi H_{\mathrm{ext}}^2 +2K+\frac{m^2H^2_{\mathrm{ext}}}{\chi H_{\mathrm{ext}}^2 +2K}\bigr].
\end{equation}

\begin{figure}[h] 
	\centering
	\includegraphics[width=0.9\linewidth]{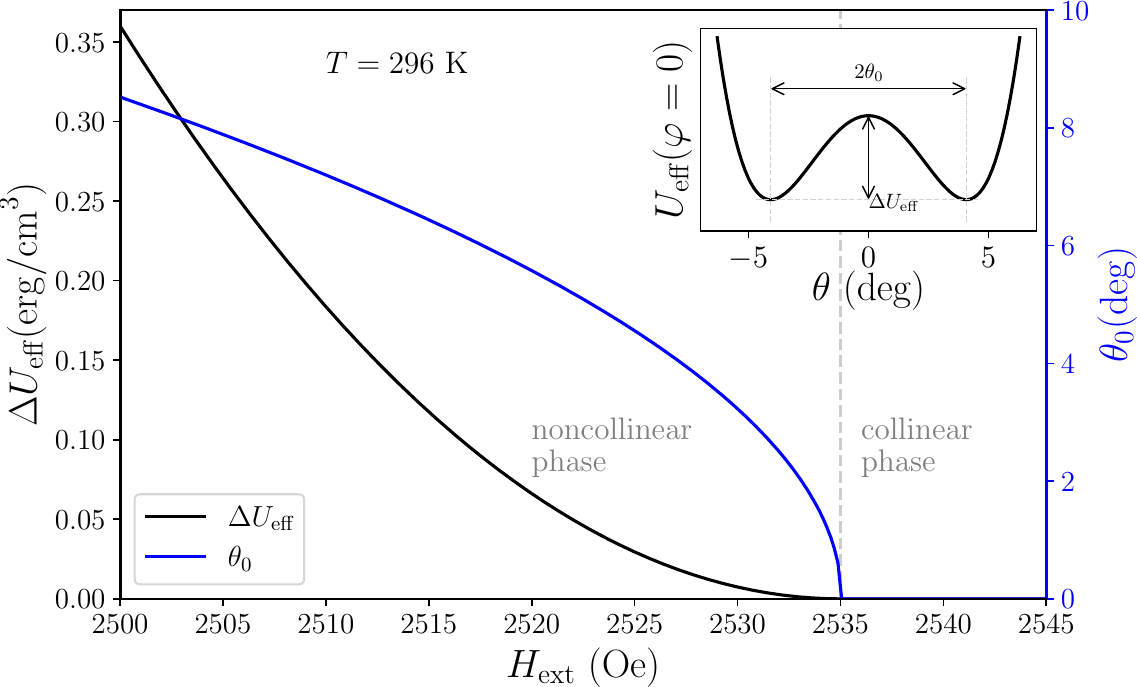}
	\caption{\justifying{
			Dependence of the initial angle $\theta_0$ and the height of the potential barrier $\Delta U_{\mathrm{eff}}$ on the external magnetic field $H_{\mathrm{ext}}$. The inset depicts the profile $U_{\mathrm{eff}}(\theta)$ for $\phi=0$, the introduced notations of $\Delta U_{\mathrm{eff}}$ and $\theta_0$ are presented.
	}}
	\label{figure1}
\end{figure}

\begin{figure*}[htb!] 
	\centering
	\includegraphics[width=1\linewidth]{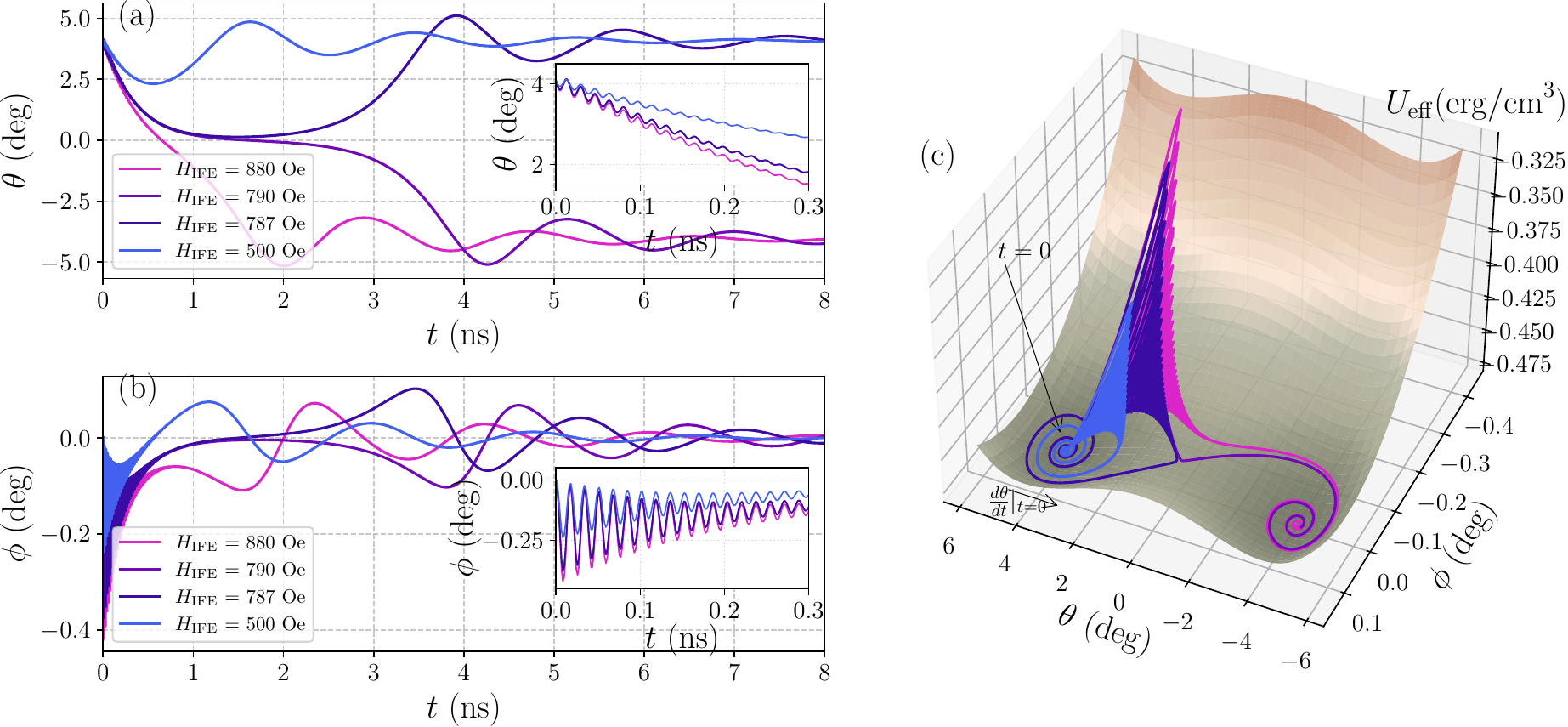}
	\caption{\justifying{Dynamics of magnetization and effective potential landscape. 
			Temporal evolution of $\theta(t)$ and $\phi(t)$ for different values of $H_{\mathrm{IFE}} = 500, 787, 790, 880$~Oe acting on the system, (a) and (b), respectively. Insets display the same $\theta(t)$ and $\phi(t)$ but at a different time range.
			(c) 3D representation of the effective potential $U_{\mathrm{eff}}(\theta,\phi)$ with superimposed dynamical trajectories $\phi(t)$ and $\theta(t)$. 
	}}
	\label{figure2}
\end{figure*}

Figure~\ref{figure1} illustrates two minima $U_\mathrm{eff}(\theta, \phi)$ function corresponding to the two equilibrium states $\pm \theta_0$ and the barrier between them $\Delta U_{\mathrm{eff}}$. Numerical calculations are performed for material parameters typical for similar iron-garnets with compensation point like $\mathrm{(BiLu)}_3(\mathrm{Fe}\mathrm{Ga})_5\mathrm{O}_{12}$ ~\cite{krichevsky2023unconventional, ignatyeva2025high}: $T_m=333$K, we work at the temperature of $T=296$~K, $H_{\mathrm{ext}} = 2527$~Oe, $K = 2660$~erg/cm$^3$, {$\chi$}$= 9.4 \times 10^{-5}$, the Gilbert damping parameter $\alpha = 10^{-4}$, $m = 2.33$~emu/cm$^3$, $m/M = 6\cdot 10^{-3}$ (Supplementary).

Thus, in RIG film in the non-collinear phase there are two degenerate states with opposite signs of $\theta_0$ angles. For realization of data writing and processing, these states can be treated as {``$0_w$'' (the potential well corresponding to $\theta_0>0$) and ``$1_w$'' ($\theta_0<0$), correspondingly.}

Magnetization dynamics are governed by the nonlinear equations of motion for angular variables $\theta(t)$ and $\phi(t)$ derived from the Lagrangian (Supplementary, Eqs.~\eqref{eq:theta}--\eqref{eq:varphi}). Unlike previous linear approximations \cite{krichevsky2023unconventional, ignatyeva2025high}, these equations account for the complex potential characterized by two minima.

Figure~\ref{figure2} illustrates the dynamics triggered by a $\sigma^+$ femtosecond pulse with right-handed circular polarization via the inverse Faraday effect (IFE). Optomagnetic influence of the circularly polarized light on spins can be described in terms of an effective magnetic field $\mathbf{H}_{\mathrm{IFE}} \propto \mathrm{Re}(a^*)[\mathbf{E} \times \mathbf{E}^*]$ \cite{gribova2026unified} oriented along the light k-vector. The field $\mathbf{H}_{\mathrm{IFE}}$ is directed either parallel or antiparallel to the light $\mathbf{k}$-vector depending on the light helicity, magnetization direction and sign of the magneooptical parameter $a$. For example, for the Bi-substituted RIG at the wavelength of 800 nm $\mathrm{Re}(a^*)<0$, therefore, for the $\sigma_+$ helicity $\mathbf{H}_{\mathrm{IFE}}$ is directed opposite to $\mathbf{k}$. Given the pulse duration $\tau = 0.05$--$0.2$~ps, the excitation is impulsive. Numerical solutions are obtained using initial conditions $\theta(0) = \theta_0$, $\phi(0) = 0$, $\left.\frac{d\theta}{dt}\right|_{t=0} = \gamma^2  (H_{\mathrm{ext}}\cos(2\theta_0)-\frac{m}{\chi}\cos\theta_0) \tau H_{\mathrm{IFE}}$, and $\left.\frac{d\phi}{dt}\right|_{t=0} = 0$. 
Temporal evolutions of $\theta(t)$ and $\phi(t)$ (Fig.~\ref{figure2}(a,b)) are presented for four $H_{\mathrm{IFE}}$ values, where $H_{\mathrm{IFE}}=1000$~Oe corresponds to a fluence $J=4.4$~mJ/cm$^2$ \cite{ignatyeva2025high}.

Figure~\ref{figure2}(c) visualizes the magnetization trajectories in a three-dimensional plot of the effective potential $U_{\mathrm{eff}}(\theta, \phi)$. The numerically calculated paths, $U_{\mathrm{eff}}(\theta(t), \phi(t))$, correspond to excitation by circularly polarized pulses with ${H}_{\mathrm{IFE}} = 500, 787, 790, \text{ and } 880$~Oe.

The precession amplitude grows with optical pulse energy. At low excitation (${H}_{\mathrm{IFE}} = 500$~Oe), the system undergoes small-angle oscillations around the initial equilibrium. Increased pulse energy yields larger amplitudes (${H}_{\mathrm{IFE}} = 787$~Oe) until the potential barrier is overcome (${H}_{\mathrm{IFE}} \geq 790$~Oe). This results in magnetization switching from the initial state at $\theta_0 = 4.1^{\circ}$ to the opposite equilibrium at $-\theta_0$.

This defines a threshold $H_{\mathrm{IFE}}^{\sigma_{\pm}}$ for switching, depicted in Fig.~\ref{figure3} (red for $H_{\mathrm{IFE}}^{\sigma_{+}}$, blue for $H_{\mathrm{IFE}}^{\sigma_{-}}$). The corresponding pulse fluences, estimated using parameters from Ref.~\cite{ignatyeva2025high}, are indicated on the right axis of Fig.~\ref{figure3}.

\begin{figure}[h] 
    \centering
    \includegraphics[width=1\linewidth]{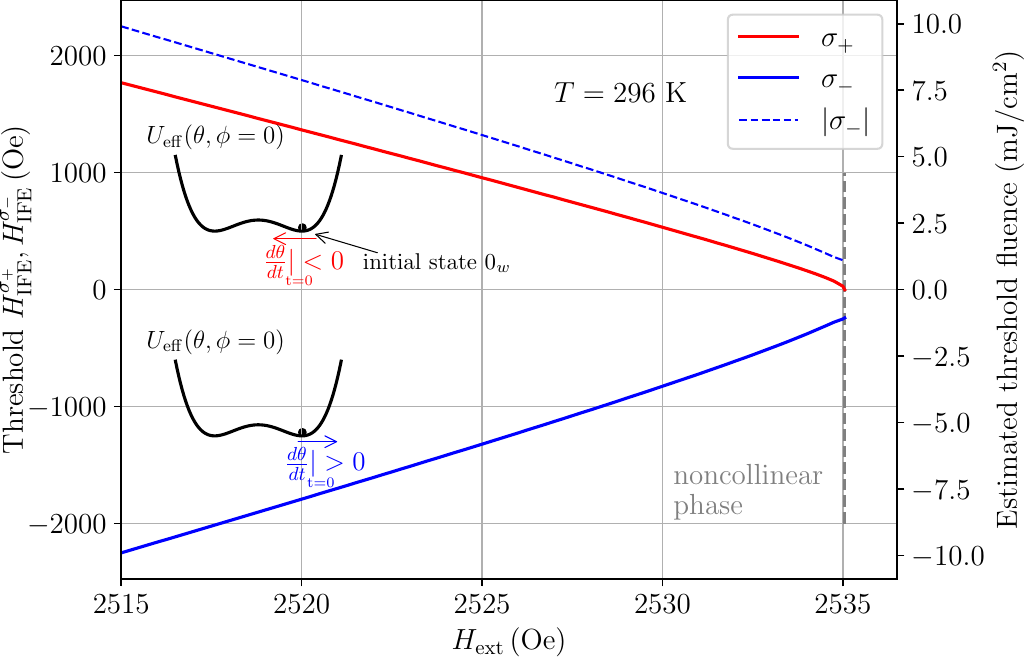}
    \caption{\justifying{Dependence of the minimal effective inverse Faraday effect field $H_{\mathrm{IFE}}$ (left axis) and optical pump fluence (right axis) required for the magnetization switching on the applied in-plane external magnetic field $H_{\mathrm{ext}}$. Two curves are shown: one corresponding to the case with negative initial angular velocity $\frac{d\theta}{dt}|_{t=0} < 0$ (right circular polarization $\sigma_{+}$), and another for $\frac{d\theta}{dt}|_{t=0} > 0$ (left circular polarization $\sigma_{-}$), where switching is less efficient.
}}
    \label{figure3}
\end{figure}

Figure~\ref{figure3} illustrates the dependence of the minimum effective inverse Faraday effect (IFE) field required for the magnetization switching on the applied in-plane field $H_{\mathrm{ext}}$. Variations in $H_{\mathrm{ext}}$ shift the system's position on the phase diagram, altering the effective potential and the initial equilibrium $\theta_0$, which consequently modifies the threshold field.

A distinct asymmetry in switching thresholds is observed between $\sigma_+$ and $\sigma_-$ helicities. For an initial state $\theta_0 > 0$, the $\sigma_+$ helicity yields a lower threshold ($H_{\mathrm{IFE}}^{\min}$, $J_{\min}$) because the initial angular velocity $\dot{\theta}|_{t=0} < 0$ favors motion towards the potential barrier. Conversely, $\sigma_-$ pulses ($\dot{\theta}|_{t=0} > 0$) initially push the system away from the barrier; Gilbert damping then causes energy loss before the system reverses direction, resulting in a higher threshold ($H_{\mathrm{IFE}}^{\max}$, $J_{\max}$). This relationship is inverted if the initial state is $\theta_0 < 0$.

This threshold disparity enables deterministic bit writing. External magnetic field $H_{\text{ext}}$ and corresponding equilibrium angle $\theta_0$ (Fig.~\ref{figure1}) define the system's initial state. By selecting a pulse energy $J$ such that $J_{\min} < J < J_{\max}$, the final magnetization state is governed solely by helicity: a $\sigma_+$ pulse resets the system to $-\theta_0$ (bit "$1_w$") and a $\sigma_-$ pulse to $\theta_0$ (bit "$0_w$"), independent on the initial orientation. This mechanism provides a robust pathway for the non-absorptive optical data storage in the non-collinear phase of ferrimagnets.

At finite temperature, there is a finite probability for the magnetization to flip and reverse its direction. The mean time between two flips is called the N\`eel relaxation time $\tau_N$ and is given by the Néel-Arrhenius equation  \cite{neel1949theorie,li2004thermally,bance2015thermal} 
\begin{equation}
	\tau_N = \tau_0 \exp\Bigl[\Delta U_{\mathrm{eff}} V/(kT)\Bigr],
\end{equation}
where $V$ {is the volume of remagnetized area} and $1/\tau_0$ is an attempt frequency \cite{fiedler2012direct}, {which describes the probability of magnetization reversal and has the typical values of} {$10^{9}$~Hz}~\cite{bance2015thermal,li2004thermally}. {For the configuration used above and magnetic bit size $10\times 10\times1.8\mu$m ($V=1.8\cdot10^{-10}$~cm$^3$), $\Delta U_{eff} = 0.0075$erg/cm$^3$ and $T=296$~K. The relation $\frac{\Delta U_{\mathrm{eff}}V}{kT} = 33$ holds, and, therefore, $\tau_N\approx3$ days, which is sufficiently large for logic operations.}
Stable smaller bits are available if higher potential barriers are used, i.e. if one adjust the magetic state further apart from the transition boarder between the collinear and non-collinera magnetic phases at the phase diagramm by either varying  temperature or the external magnetic field. 

In summary, we have theoretically and numerically investigated nonlinear {spin dynamics and optically-induced magnetization switching in a bistable rare-earth iron garnet with the magnetization compensation point. A key feature is that the magnetization switching is launched without any light absorption via the optomagnonetic effect, the inverse Faraday effect, namely. It exhibits a threshold of the effective magnetic field, i.e. pulse fluence which can provide the magnetization switching. The threshold depends on the applied external magnetic field and the helicity of the incident optical pulse. It gives a perspective for the non-absorptive light impact on spins to deterministically put the system into the desired magnetization state, and to write a magnetic bit. The dependence of the switching threshold on the optical pulse helicity and magnetic phase of the sample opens new possibilities for construction of the optomagnonic logic elements.   These results highlight the potential of rare-earth iron garnets as a material platform for fast, energy-efficient and reconfigurable optomagnonic logics, and provoke future experimental implementations of the optomagnonic computing devices.}

{
This work was financially supported by Russian Science Foundation (Project No. 23-62-10024) in part related to electromagnetic modeling of the spin dynamics. N.I.G. and V.I.B. also acknowledge support from the Foundation for the Advancement of Theoretical Physics and Mathematics “BASIS” (Project No. 25-1-1-49-4) for analytical theoretical studies.}

\putbib[main.bib]
\end{bibunit}


\clearpage
\newpage
\maketitle
 \onecolumngrid                   
\begin{bibunit}                                                                 

\setcounter{section}{0}
\setcounter{equation}{0}
\setcounter{figure}{0}
\numberwithin{equation}{section}
\renewcommand\thesection{S.\arabic{section}}
\renewcommand\theequation{S.\arabic{equation}}
\renewcommand\thefigure{S.\arabic{figure}}
\section*{Supplementary}\label{appendix1}

The Lagrangian formalism is employed to describe the dynamics of the sublattice magnetization vectors and derive the equations of motion for the antiferromagnetic vector $\mathbf{L}$, parameterized by spherical angles $\theta$ and $\phi$ as $\mathbf{L}=L(\cos\theta\cos\phi,\cos\theta\sin\phi, \sin\theta)$. The two-sublattice system is characterized by $\mathrm{Fe}^{3+}$ ions on  octahedral ($\mathbf{M}_a$) and tetrahedral ($\mathbf{M}_d$) magnetization vectors, with magnitudes $M_a$ and $M_d$, and spherical angles $\theta_a, \phi_a$ and $\theta_d, \phi_d$, respectively ($\mathbf{M}_{i}=M_{{i}}(\cos\theta_{{i}}\cos\phi_{{i}}, \cos\theta_{{i}}\sin\phi_{{i}}, \sin\theta_{{i}})$ with $i = a,d$).

The angles of antiferromagnetic vector and two-sublattices are related by 
\begin{align}
    &\theta_{d}=\theta+\varepsilon, \qquad \theta_a=-\theta+\varepsilon,\\
    &\phi_{d}=\phi+\beta,\qquad \phi_a=\pi+\phi-\beta,
\end{align}
where the parameters $\varepsilon\ll1$ and $\beta\ll1$ characterize the noncollinearity of the magnetization vectors 
of the sublattices \cite{krichevsky2023unconventional}. The Lagrangian of the system is given by~\cite{davydova2019ultrafast, zvezdin1979dynamics, blank2021thz,krichevsky2023unconventional} and rewritten using $\theta$, $\varepsilon$, $\phi$, $\beta$:
\begin{align}
    \mathcal{L} &= - \frac{m}{\gamma} \sin\theta \, \dot{\phi} - \frac{M}{\gamma} \left(\epsilon \dot{\phi} - \beta \dot{\theta}\right)\cos\theta - \Phi,\label{supp:lagrangian}\\
    \Phi &= -(\mathbf{M}_a+\mathbf{M}_d)\mathbf{H}_{\mathrm{ext}}+\Lambda \mathbf{M}_a\mathbf{M}_d-K_a\frac{(\mathbf{M}_a\mathbf{n})^2}{{M}_a^2}-K_d\frac{(\mathbf{M}_d\mathbf{n})^2}{{M}_d^2}\\
    &\approx -mH_{\mathrm{ext}}\cos\theta\cos\phi+M H_{\mathrm{ext}}(\beta \cos\theta\sin\phi+\epsilon\sin\theta\cos\phi)+\Lambda M_a M_d(\epsilon^2+\beta^2\cos^2\theta-1)-K\sin^2\theta \label{supp:potential}
\end{align}
where $\gamma=\gamma_d=\gamma_a$ is the gyromagnetic ratios assumed equal for both sublattices, $\Phi$ is the potential energy \cite{krichevsky2023unconventional,belov1976spin}, $K$ is the effective magnetic {uniaxial} anisotropy constant calculated as the sum of sublattice anisotropies $K_{\mathrm{Fe}}$ and $K_{\mathrm{R}}$: $K = K_{\mathrm{a}}+K_{\mathrm{d}}$, $M = M_d + M_a$ is the sum of the sublattice magnetizations, $m = M_d - M_a$ is the difference between the sublattice magnetizations, $\Lambda>0$ is the intersublattice exchange constant, that represents the antiferromagnetic character of the exchange interaction between the sublattices. The effect of an external magnetic field $H_{\mathrm{ext}}$ is also considered in potential energy, under the assumption that the applied field is much weaker than the exchange field.

Next, substituting the relation for the potential energy (Eq.~\ref{supp:potential}) into the Lagrange function (Eq.~\ref{supp:lagrangian}) and
considering $(\epsilon,\beta)$ as a generalized coordinates set, with respect to them one can calculate the Lagrange equations that take the form
\begin{align}
    \frac{2\Lambda M_a M_d}{M}\epsilon=-\frac{\dot{\phi}}{\gamma}\cos\theta -H_{\mathrm{ext}}\sin\theta \cos\phi\label{supp:el1}\\
    \frac{2\Lambda M_a M_d}{M}\beta\cos\theta =\frac{\dot{\theta}}{\gamma}-H_{\mathrm{ext}}\sin\phi\label{supp:el2}
\end{align}

Substituting Eqs.~(\ref{supp:el1,supp:el2}) into Eqs.~\ref{supp:lagrangian} one can exclude the
noncollinearity parameters $\epsilon$ and $\beta$ and reduce the Lagrangian of the two-sublattice system to a
function of only two angles $\theta$ and $\phi$ of the antiferromagnetic vector $\mathbf{L}$ and their time derivatives. The effective (reduced) Lagrangian is presented in \cite{krichevsky2023unconventional} and takes the form
\begin{equation}
    \mathcal{L}_{\mathrm{eff}} = \frac{\chi}{2}\Bigl[
    \Bigl(
\frac{\dot{\phi}}{\gamma}\cos\theta+H_{\mathrm{ext}}\sin\theta\cos\phi
    \Bigr)^2 
    +
    \Bigl(
H_{\mathrm{ext}}\sin\phi -\frac{\dot{\theta}}{\gamma}
    \Bigr)^2 
    \Bigr]
    -\dot{\phi}\frac{m}{\gamma}\sin\theta + mH_{\mathrm{ext}}\cos\theta\cos\phi+K\sin^2\theta.\label{supp:lagrangian_eff}
\end{equation}

\begin{figure}[h] 
	\centering
	\includegraphics[width=0.5\linewidth]{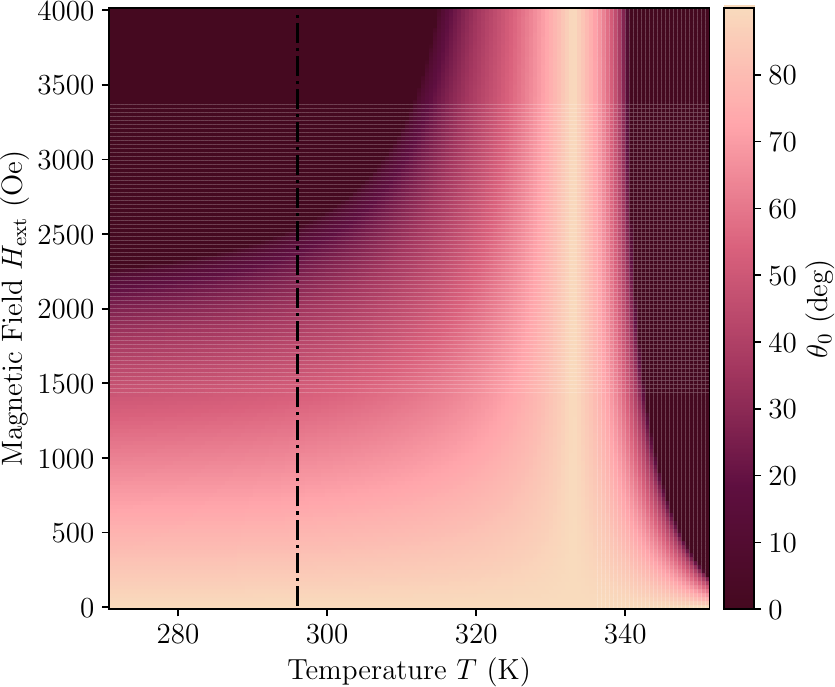}
\caption{\justifying{$\theta_0(H_{\mathrm{ext}}, T)$ phase diagram using the parameters from Section II.
	}}
	\label{supp:diagram}
\end{figure}

{The equilibrium states of the two-sublattice system are found as the minimum of the effective 
potential energy from the effective Lagrangian (Eq.~\ref{supp:lagrangian_eff}) obtained using the known relationship between the Lagrange and the Hamiltonian 
functions for the stationary case $\dot{\theta}=0$ and $\dot{\phi}=0$: 
\begin{equation}
    U_{\mathrm{eff}} = -\frac{\chi}{2} H_{\mathrm{ext}}^2 (\sin^2\theta \cos^2\phi+\sin^2\phi)
	-  m H_{\mathrm{ext}}\cos\theta\cos\phi - K\sin^2\theta.\label{effective_u}
\end{equation}
}
Eq.~\ref{effective_u} shows that antiferromagnetic vector L in the equilibrium state characterized by $\theta_0$ and $\phi_0$ angles.  In the $H_{ext}-T$ phase diagram $\theta_0(H_{ext}, T)$ of a ferrimagnet two main phases are possible: collinear and non-collinear ones \cite{krichevsky2023unconventional}, that are presented in Fig.~\ref{supp:diagram}.

{In order to derive the Euler–Lagrange equations for the angles $\theta$ and $\phi$ describing the antiferromagnetic vector $\mathbf{L}$ \cite{krichevsky2023unconventional}, we employ the Lagrangian and Rayleigh dissipation functions\cite{krichevsky2023unconventional}. This approach allows us to obtain a system of second-order differential equations governing the dynamics of these two variables }
\begin{align}
	-\frac{M^2\ddot{\theta}}{\delta\gamma^2}+&\frac{M^2 H_{\mathrm{ext}}}{\delta\gamma}\dot{\phi}\cos\phi - \frac{M^2}{\delta}\Bigl(\frac{\dot{\phi}^2}{\gamma^2}\sin\theta\cos\theta-
	H_{\mathrm{ext}}^2\sin\theta\cos\theta\cos^2\phi
	-\frac{\dot{\phi}}{\gamma}H_{\mathrm{ext}}\cos\phi\cos2\theta\Bigr)-\nonumber
	\\
	&\dot{\phi}\frac{m}{\gamma}\cos\theta-mH_{\mathrm{ext}}\sin\theta\cos\phi+K\sin2\theta=\frac{\alpha M}{\gamma}\dot{\theta}\label{eq:theta}
\end{align}
\begin{align}
	\frac{M^2}{\delta\gamma}\Bigl(-\frac{\dot{\phi}}{\gamma}&\dot{\theta}\sin 2\theta + H_{\mathrm{ext}}\dot{\theta}\cos 2 \theta\cos\phi-H_{\mathrm{ext}}\dot{\phi}\sin\theta
	\cos\theta\sin\phi+\frac{\ddot{\phi}}{\gamma}\cos^2\theta\Bigr) -\frac{m}{\gamma}\dot{\theta}\cos\theta+\nonumber\\
	&\frac{M^2}{\delta}\Bigl(\frac{\dot{\phi}}{\gamma}H_{\mathrm{ext}}\sin\theta\cos\theta\sin\phi +\frac{\dot{\theta}}{\gamma}H_{\mathrm{ext}}\cos\phi+
	H_{\mathrm{ext}}^2\sin^2\theta\cos\phi\sin\phi-H_{\mathrm{ext}}^2\cos\phi\sin\phi\Bigr)+\nonumber\\
	&mH_{\mathrm{ext}}\cos\theta\sin\phi=-\frac{\alpha M}{\gamma}\dot{\phi}\cos^2\theta\label{eq:varphi}
\end{align}
{where $\alpha$ is the Hilbert damping parameter, $\delta=2\Lambda M_{\mathrm{Fe}} M_{\mathrm{R}}$ is the difference between the sublattice magnetizations, $\Lambda>0$ is the 
intersublattice exchange constant and the “+” sign of the second term represents the antiferromagnetic character of the exchange interaction between the sublattices. These equations describe the dynamics of the system in the presence of damping and external forces, and are valid within the quasi-antiferromagnetic approximation.}

\putbib[main.bib]
\end{bibunit}
\end{document}